\documentclass[aoas,MSNbibl,nameyear,dvips]{arximspdf}
\usepackage{dcolumn}
\usepackage{graphicx}

\doi{10.1214/14-AOAS727} 
\volume{8}
\issue{2}
\pubyear{2014}
\firstpage{703}
\lastpage{725}

\makeatletter
\newcolumntype{d}[1]{D{.}{.}{#1}}
\makeatother

\begin{document}
\begin{frontmatter}

\title{Effect of breastfeeding on gastrointestinal \hspace*{-4pt}infection in
infants: A targeted maximum likelihood approach for clustered longitudinal data}
\runtitle{The effect of breastfeeding on GI using clustered TMLE}

\begin{aug}
\author[a]{\fnms{Mireille~E.}~\snm{Schnitzer}\corref{}\thanksref{t1,m1}\ead[label=e1]{mireille.schnitzer@umontreal.ca}},
\author[b]{\fnms{Mark~J.}~\snm{van~der~Laan}\thanksref{t2,m2}\ead[label=e2]{laan@berkeley.edu}},
\author[c]{\fnms{Erica~E.~M.}~\snm{Moodie}\thanksref{t3,m3}\ead[label=e3]{erica.moodie@mcgill.ca}}
\and
\author[c]{\fnms{Robert~W.}~\snm{Platt}\thanksref{t4,m3}\ead[label=e4]{robert.platt@mcgill.ca}}
\runauthor{Schnitzer, van~der~Laan, Moodie and Platt}
\thankstext{t1}{Supported by an Alexander Graham Bell Canada Graduate
Scholarship from the Natural Sciences and Engineering Research Council
of Canada (NSERC) and a Bourse de stage internationale from the Fonds
de recherches du Qu\'ebec en nature et technologies.}
\thankstext{t2}{Supported by the National Institutes of Health (NIH)
under Targeted Empirical Super Learning in HIV Research NIH Grant \#5R01AI74345.}
\thankstext{t3}{Supported by a University Faculty Award from NSERC.}
\thankstext{t4}{Supported by a Chercheur-boursier award from the Fonds
de la recherche en sant\'e du Qu\'ebec.}
\affiliation{Universit\'e de Montr\'eal\thanksmark{m1}, University of
California, Berkeley\thanksmark{m2}\break and McGill University\thanksmark{m3}}
\address[a]{M. E. Schnitzer\\
Facult\'e de pharmacie\\
Universit\'e de Montr\'eal\\
Montr\'eal, Qu\'ebec H3T 1J4\\
Canada\\
\printead{e1}}

\address[b]{M. J. van der Laan\\
Division of Biostatistics\\
University of California, Berkeley\\
Berkeley, California 9470-7358\\
USA\\
\printead{e2}}

\address[c]{E. E. M. Moodie\\
R. W. Platt\\
Department of Epidemiology,\\
\quad Biostatistics and Occupational Health\\
McGill University\\
Montr\'eal, Qu\'ebec H3A 1A2\\
Canada\\
\printead{e3}\\
\phantom{E-mail:\ }\printead*{e4}}

\end{aug}

\received{\smonth{8} \syear{2013}}
\revised{\smonth{1} \syear{2014}}

%
\begin{abstract}
{The PROmotion of Breastfeeding Intervention Trial (PROBIT)
cluster-randomized a program encouraging breastfeeding to new mothers
in hospital centers. The original studies indicated that this
intervention successfully increased duration of breastfeeding and
lowered rates of gastrointestinal tract infections in newborns.
Additional scientific and popular interest lies in determining the
causal effect of longer breastfeeding on gastrointestinal infection.
In this study, we estimate the expected infection count under various
lengths of breastfeeding in order to estimate the effect of
breastfeeding duration on infection. Due to the presence of baseline
and time-dependent confounding, specialized ``causal'' estimation
methods are required. We demonstrate the double-robust method of
Targeted Maximum Likelihood Estimation (TMLE) in the context of this
application and review some related methods and the adjustments
required to account for clustering. We compare TMLE (implemented both
parametrically and using a data-adaptive algorithm) to other causal
methods for this example.
In addition, we conduct a simulation study to determine (1) the
effectiveness of controlling for clustering indicators when
cluster-specific confounders are unmeasured and (2) the importance of
using data-adaptive TMLE.}
\end{abstract}

%

\end{frontmatter}

\section{Introduction}\label{sec1}

The PROmotion of Breastfeeding Intervention Trial (PROBIT)
[Kramer et al. (\citeyear{Kramer:001,Kramer:002})] was undertaken in order to obtain randomized
control trial evidence of the health effects of longer breastfeeding.
This was done by cluster randomizing a breastfeeding support
intervention which encouraged exclusivity and duration. The effect of
the PROBIT intervention on gastrointestinal tract infection in the
newborns was originally evaluated using a stratified intention-to-treat
analysis. The results indicated a significant reduction in infection
incidence for infants whose mothers had been assigned to the
intervention group [\citet{Kramer:001}]. The intervention was presumably
effective because it successfully encouraged breastfeeding, which
subsequently improved infant health. However, because breastfeeding
itself was not randomized, the estimated effect obtained in the study
can at best be considered a biased assessment of the effect of
breastfeeding on infection. Due to the ethical and practical
impossibility of randomizing breastfeeding, estimation of the causal
effect of breastfeeding must be obtained through statistical methods.

Our goal is therefore to estimate the causal effect of breastfeeding
duration on the number of infections a newborn is expected to
experience in their first year. One of the challenges involved in
analyzing this effect is the confounding presence of intermediate
infections (occurring at any time during the year). The presence of an
infection affects both the continuation of breastfeeding and the
outcome (since it deterministically increases the outcome by one).
Therefore, intermediate infection is a \textit{time-dependent
confounder}. Since infection is also hypothesized to be affected by
previous breastfeeding status, standard regression methods (including
or excluding the time-dependent confounder) may produce a biased
estimate of the causal parameter [\citet{Robins:002}]. Causal methods
are therefore required to isolate the desired effect. Additional
confounding also occurs due to baseline differences in the study group
and by informative participant dropout.

Many longitudinal methods have been developed that correctly take into
account time-dependent confounders predicted by past exposure. One such
method is inverse probability of treatment weighting (IPTW) for
marginal structural models [\citet{Hernan:001,Robins:001}]. However,
IPTW is not semiparametric efficient\break [\citet{Robins:005}] and has poor
performance under certain common scenarios~[\citet{Petersen:002}]. The
shortcomings of simple weighting methods have since spurred the
development of new estimators with better properties. Efficient
estimating equation methodology [\citet{Robins:005,vdl:005,Bang:001}]
produces estimators that are double robust (consistent under partial
model misspecification) and efficient when correctly specified.
Targeted maximum likelihood estimation (TMLE) [\citet{vdl:001}] shares
these properties, but because it is a substitution estimator, it can be
made to be stable and produce estimates bounded within the parameter
space in some situations where IPTW performs poorly~[\citet
{Gruber:001}]. In addition, TMLE is often implemented fully
nonparametrically, which avoids modeling errors caused by incorrect
parametric assumptions.

\citet{vdl:002} established a TMLE procedure for longitudinal data
based on a binary decomposition of the intermediate variables (the
time-dependent confounders). This method has been described and
implemented by~\citet{Rosenblum:002} and \citet{Schnitzer:001} for two
time points, and~\citet{Stitelman:001} for a survival outcome. However,
the implementation of this method for large numbers of time points
results in heavy computational requirements and a restriction on the
form of the data (specifically, requiring discretized intermediate
covariates). More recently, \citet{vdl:004} developed a simpler and
more flexible implementation of TMLE for longitudinal data based on the
ideas of~\citet{Bang:001}.

An initial causal analysis of the PROBIT study using different
double-robust causal methods was performed by~\citet{Schnitzer:001} but
was limited to two time points. In this paper, after giving more
details about the PROBIT study and the scientific question of interest
(Section~\ref{sec2}), we describe several options for potentially unbiased
estimation of the effect of breastfeeding on infection: (a)~G-computation~[\citet{Robins:002}], (b) a variant of G-computation that
we call sequential G-computation~[\citet{Bang:001}], and (c) a
longitudinal TMLE based on sequential G-computation~[\citet{vdl:004}]
(Section~\ref{sec3}). The subsection on the longitudinal TMLE demonstrates a 6
time-point implementation for estimation of the effect of breastfeeding
duration on gastrointestinal tract infection, with modified variance
estimation reflecting the clustered design of the PROBIT. In Section~\ref{sec4}
we present the results of analyzing the PROBIT data with each of these
methods in addition to IPTW. Finally, we compare this TMLE approach to
the other causal techniques for longitudinal data in a simulation study
designed to imitate the analysis of the PROBIT data.

\section{The PROBIT data}\label{sec2}
The PROBIT study paired participating maternal hospitals according to
(1) geographic region in Belarus, (2) urban or rural status, (3)~number
of deliveries per year and (4) breastfeeding rates upon discharge. One
hospital of each pair was then assigned to receive a breastfeeding
support intervention that involved retraining all midwives, nurses and
physicians involved in labor, delivery and the postpartum hospital
stay. The control hospitals were assigned to continue their current
practice. Thirty-four hospitals were initially randomized, but three
were dropped from the study due to eventual refusal to follow the
assignment or falsification of data.

The PROBIT study enrolled healthy, full-term, singleton infants of
mothers who intended to breastfeed, weighing at least 2500~g, soon after
birth. Follow-up visits were scheduled at 1, 2, 3, 6, 9 and 12 months
of age to record various measures of health and size, including number
of gastrointestinal infections over each time interval. At each
follow-up visit, it was established whether the mother continued to breastfeed.

Within the 31 hospitals, 17,046 mother/infant pairs were recruited into
the trial. Of these, ten were missing necessary baseline information
and were removed from the analysis. The remaining 17,036 subject pairs
were used in the analysis. Characteristics of the complete data set
(including missing data summaries) are presented in Table~\ref{characteristics2}. Within the hospitals, the number of recruited
patients varied between 232 and 1180 with median 471.

\begin{table}
\caption{Characteristics at baseline of the 17,046 mother-infant pairs
in the PROBIT data set}\label{characteristics2}
\begin{tabular*}{\textwidth}{@{\extracolsep{\fill}}lccc@{}}
\hline
\textbf{Characteristic} & \multicolumn{2}{c}{\textbf{Summary}}&\textbf{N.~missing}\\
\hline
\emph{Numeric variables}&Median&IQR\tabnoteref[a]{ttl1}&\\
Age of mother (years)& 23 & $(21,27)$& \\
N. previous children&0&$(0,1)$&\\
Gestational age (months)& 40 & $(39,40)$&\\
Infant weight (kg)& 3.4&$(3.2,3.7)$&\\
Infant height (cm)&52.00&$(50.00,53.00)$&\\
Apgar score\tabnoteref[b]{ttl2}& 9 &$(8,9)$&5\\
Head circumference (cm)&35&$(34,36)$&3\\[3pt]
\emph{Binary variables}&N.&$\%$&\\
Smoked during pregnancy& 389&2.28&\\
History of allergy&750&4.40&\\
Male child&8827&52& \\
Cesarean&1974&12&\\
Mother's education&&&2\\
\quad Some high school&663&4&\\
\quad High school&5497&32&\\
\quad Some university&8568&50&\\
\quad University&2316&14&\\
Geographic region&&&\\
\quad East Belarus, urban&5615&33&\\
\quad East Belarus, rural&2706&16&\\
\quad West Belarus, urban&4380&26&\\
\quad West Belarus, rural&4343&25&\\
\hline
\end{tabular*}
\tabnotetext[a]{ttl1}{IQR: inter-quartile range.}
\tabnotetext[b]{ttl2}{The Apgar score is an assessment of newborn
health (range 1--10) where 8$+$ is vigorous, 5--7 is mildly depressed and
4$-$ is severely depressed~[\citet{Finster:001}]. A range of 5--10 was
observed in PROBIT due to entry restrictions on weight and health at baseline.}
\end{table}

Measured baseline potential confounders of the effect of breastfeeding
on infection (and predictors of outcome) were chosen to be mother's
education, mother's smoking status during pregnancy, mother's age,
family history of allergy, number of previous children, whether the
birth was by cesarean section, gender of child, gestational age, Apgar
score for health of the newborn, geographic region, and the weight,
height, head circumference at birth, and hospital. The hospital (or
cluster) was included in the set of potential confounders because the
conditions of the hospital frequented by a patient can affect both
their infant's health outcome and their decision to continue
breastfeeding. In addition, since similar patients may be clustered
within a hospital, hospital may act as a proxy for unmeasured baseline
characteristics.

The hypothetical intervention of interest for this analysis was
breastfeeding up until a given time. The binary intermediate variable
at a given time was whether or not gastrointestinal infection occurred
in the interval immediately preceding the time point. The outcome is
the total number of infections occurring up until 12 months of age.

A subject was defined as censored at the first visit where information
required in the analysis was missing. The number of censored subjects
at each time point is described in Table~\ref{counts2}. Absenteeism or
study drop-out are often dependent on subject-specific characteristics
and current health, which is why adjustment for censoring was
considered necessary.

\begin{table}
\caption{Censoring, number of infections and mothers still
breastfeeding by time point}\label{counts2}
\begin{tabular*}{\textwidth}{@{\extracolsep{\fill}}lcccccc@{}}
\hline
Time point& 1 & 2 & 3 & 4 & 5 & 6\\
Month&1&2&3&6&9&12\\[3pt]
N.~censored & 284 & 500 & 326 & 491 & 717& 139\\
Cumulative N. & 284 & 784 & 1110 & 1601 & 2318 & 2457\\
Cumulative \% & 1.66 & 4.60 & 6.52 &9.40&13.61&14.42\\[3pt]
N. with infections & 171 & 232 & 230 & 443 & 518& 408\\
N. of infections&173&235&236&472&544&439\\
N. breastfeeding & 15,392 &13,128 & 10,765 &6893&4717&--\\
\hline
\end{tabular*}
\end{table}

\begin{figure}

\includegraphics{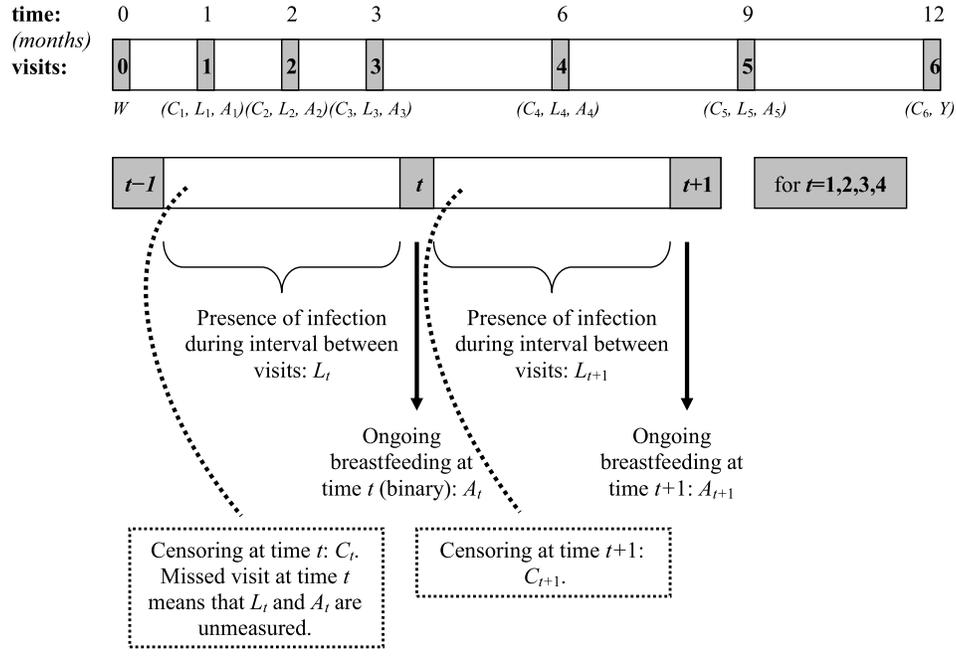}

\caption{Time-ordering of the variables in the PROBIT study. Data were
collected at baseline and six follow-up times. At each follow-up time
point, breastfeeding status ($A_t$) and presence of infection over the
past interval ($L_t$) were noted. Censoring occurring at time $t$
($C_t=1$) indicates that later breastfeeding and infection status were
not observed.}\label{figure21}
\end{figure}

At each visit, the number of gastrointestinal infections since the last
visit were counted. In addition, breastfeeding status at that time was
obtained. There is therefore uncertainty about exact time-ordering of
each infection and breastfeeding cessation within a time interval. By
defining the exposure as breastfeeding status at time-point $t$, we can
consider that this intervention point occurs after infection counts
measured over the previous interval. With six visits, and the outcome
assessed at the sixth visit, this means that only the first five
exposure nodes are considered in the analysis. However, we observe six
censoring times (occurring before each of the six follow-up times).
Figure~\ref{figure21} gives a graphic display of the time-ordering of
the observed data.

%

Intermediate infections were considered to be an important time-varying
confounder because mothers were less likely to continue breastfeeding
when their infant became ill. Therefore, even if breastfeeding has
absolutely no effect on infection, ignoring this confounding effect
would make it seem like infants who experienced infections were also
breastfed for shorter periods of time. Table~\ref{counts2} also shows a
summary of the infection counts at each time point. Few children
experienced more than one infection during a given time interval, so
the time-dependent confounder was summarized as a binary indicator of
infection. However, we used the true number of infection counts for the outcome.

\section{Estimation for longitudinal data}\label{sec3}

As in the PROBIT study, suppose we observe longitudinal information
from $n$ individuals of the form
$O=(W,C_1,L_1,A_1,C_2,L_2,\ldots,L_{K-1},A_{K-1},C_{K},Y)$. Let $K$ be the
total number of follow-up visits, and the subscripts on each variable
indicate the visit at which that variable was measured. The variable
$W$ is the collection of potentially confounding variables at baseline.
The variables $C_t,t=1,\ldots,K$, indicate whether a subject has been
censored before the $t$th time point. Intermediate infection
was represented by $L_t,t=1,\ldots,K-1$, indicating whether the infant had
any gastrointestinal infections between time-points $t-1$ and $t$. If a
subject has been censored, define their missing $L_t$ and $Y$ values to
be zero. The variables $A_t,t=1,\ldots,K-1$, denote breastfeeding status
at time-point $t$ ($A_t=1$ means continued breastfeeding). The outcome
$Y$ is the total number of infections accrued up until and including
visit $K$. 
For any time-dependent variable $X$, we will use $\bar
{X}_t=(X_1,\ldots,X_t)$ to denote the history of $X$ up to and including $X_t$.

Let $\bar{a}=(a_1,a_2,\ldots,a_{K-1})$ denote a fixed breastfeeding
regimen. For instance, breastfeeding past the first time period, then
stopping before the second would be written as $(1,0,0,\ldots,0)$. Because
breastfeeding is approximately monotone, the regimens of interest are
equivalent to a corresponding duration of breastfeeding. Following the
Neyman--Rubin model [\citet{Rubin:001}], define the counterfactual
variable $L^{\bar{a}}_t$ as the observation $L_t$ that an individual
would have had if they had followed the breastfeeding regimen $\bar{a}$
and remained uncensored. Similarly, $Y^{\bar{a}}$~is the counterfactual
number of infections that would have been observed under breastfeeding
regimen $\bar{a}$. The target of inference is the marginal mean
counterfactual outcome, denoted $\psi_{\bar{a}}=E(Y^{\bar{a}})$. The
standard causal missing data problem arises from observing each
individual under only one breastfeeding regimen.

\subsection{The G-computation method}\label{Gcomp2}

G-computation~[\citet{Robins:002,Snowden:001}] is a likelihood-based
approach to estimating a causal parameter. It is often described as a
substitution estimator because it takes a fit of the likelihood and
substitutes it into a function to get an estimate of the parameter of
interest. 
Suppose our observed data $O$ consist of $n$ independently and
identically distributed draws from a true underlying distribution
$f(O)$. This density may be decomposed corresponding to the
time-dependent structure of the data as
\begin{eqnarray*}
f(O)&=& \underbrace{Q_{Y}(Y\mid\bar{C}_K,
\bar{A}_{K-1},\bar {L}_{K_1})\prod_{t=1}^{K-1}
Q_{L_t}(L_t \mid \bar{C}_t,\bar
{A}_{t-1},\bar{L}_{t-1},W)Q_{W}(W)}_{Q}
\\
&&{}\times \underbrace{\prod_{t=1}^{K-1}
g_{A_t}( A_t \mid\bar{L}_t,
\bar{C}_{t},\bar{A}_{t-1},W) \prod
_{t=1}^{K} g_{C_t}(C_t \mid
\bar{A}_{t-1},\bar{L}_{t-1},\bar {C}_{t-1},W)}_{g},
\end{eqnarray*}
where $Q$ is the joint conditional distribution of the $Y$, $L_t$ and
$W$ variables that can be decomposed into conditional distributions
$Q_{Y}$, $Q_{L_t},t=1,\ldots,K$, and $Q_{W}$. Similarly, $g$ is the
conditional distribution of the exposure and censoring variables that
can be decomposed into $g_{A_t},t=1,\ldots,K-1$, and $g_{C_t},t=1,\ldots,K$.

Given a fixed breastfeeding regimen, $\bar{a}$, we can define the
distribution $Q^{\bar{a}}$ of the corresponding counterfactual
variables $Y^{\bar{a}}, \bar{L}^{\bar{a}}_K, W$ (under the causal
assumptions of consistency and sequential ignorability discussed in
Section~\ref{causalassumptions}) as
\begin{eqnarray*}
Q^{\bar{a}}\bigl(Y^{\bar{a}}, \bar{L}^{\bar{a}}_K, W
\bigr) &=& Q_Y(Y\mid\bar {C}_{K}=0, \bar{A}_{K-1}=
\bar{a}_{K-1}, \bar{L}_{K-1},W)
\\
&&{}\times\prod_{t=1}^{K-1} Q_{L_t}(L_t
\mid \bar{C}_{t}=0, \bar{A}_{t-1}=\bar {a}_{t-1},
\bar{L}_{t-1},W)Q_{W}(W),
\end{eqnarray*}
where $\bar{a}_t=(a_1,\ldots,a_t)$ is the component of the fixed regime up
until time-point~$t$. The targeted parameter of interest, specifically
the marginal mean under a fixed breastfeeding regimen $\bar{a}$, can
then be described as $\hat{\psi}_{\bar{a}}=E_{Q} Y^{\bar{a}}$ where the
expectation is taken under $Q^{\bar{a}}$.

Because the intermediate variables $L_t, 1\leq t \leq K-1$, are binary,
the expression for $\psi_{\bar{a}}=E_{Q} Y^{\bar{a}}$ simplifies to
%
\begin{eqnarray}\label{Gcompform}
\psi_{\bar{a}}&=&\int_{W} \sum
_{l_1=\{0,1\}}\cdots\sum_{l_{K-1}=\{0,1\}
} E(Y \mid
C_K=0, \bar{A}_{K-1}=\bar{a}, \bar{L}_{K-1}=\bar
{l}_{K-1},W)\nonumber\\[-1pt]
&&\hspace*{102pt}{}\times \operatorname{Pr}(L_{K-1}=l_{K-1} \mid\bar{C}_{K-1}=0,
\nonumber
\\[-10pt]
\\[-10pt]
\nonumber
&&\hspace*{130pt}{}\bar{A}_{K-2}=\bar {a}_{K-2},\bar{L}_{K-2}=
\bar{l}_{K-2},W)\cdots
\\[-1pt]
&&\hspace*{102pt}{}\times \operatorname{Pr}(L_1=l_1 \mid C_1=0,W)Q_{W}(W)\,dW.
\nonumber
\end{eqnarray}
Each component of the above expression can be estimated from the
observed data. Only the conditional mean of $Y$ and the conditional
probabilities for $L_t, 1\leq t \leq K$, must be fit to produce a
G-computation estimate. The mean and the conditional probabilities can
be estimated using any parametric method as desired.

To obtain an estimate of the parameter using G-computation, first get a
prediction of each conditional expectation and probability in equation
(\ref{Gcompform}) for each subject, $i$. The $Q_{W}$ can be estimated
using the empirical density so that $Q_{W}(w_i)=1/n$ for each subject
(with baseline variables $w_i$). Then, the predicted values for the
conditional expectation and probabilities are combined according to
equation~(\ref{Gcompform}), where the integral is replaced by summation
over all subjects, $i$.

G-computation does not rely on the full specification of the density
$Q$. However, it requires correct specification of the conditional
models for the mean and each of the probabilities in order to obtain
unbiased estimation of the parameter $\psi_{\bar{a}}$. No closed form
or asymptotic result is available for the G-computation standard error,
so using a nonparametric bootstrap is often suggested~[\citet
{Snowden:001}]. To properly assess the variance in the clustered
design, the analyst might use the pairs clustered bootstrap~[\citet
{Cameron:001}] by resampling clusters instead of individuals.

\subsection{Sequential G-computation formulation}\label{altG}

As suggested by Bang and Robins (\citeyear{Bang:001}) and used by~\citet{vdl:004}, an
alternative decomposition of the parameter of interest, and therefore
an alternative to the standard likelihood G-computation, can be
constructed by taking sequential expectations of the outcome. Their
result is an application of the property of iterated expectations.

Under the causal assumptions of sequential exchangeability and
consistency, the marginal mean under breastfeeding regime $\bar{a}$ and
no censoring can be reexpressed as
\begin{eqnarray}
\psi_{\bar{a}}&=&E\bigl(Y^{\bar{a}}\bigr)
\nonumber
\\[-1pt]
&=&E\bigl\{ E( Y \mid C_K=0,\bar{A}_{K-1}=
\bar{a}_{K-1},\bar{L}_{K-1},W ) \bigr\}
\nonumber
\\[-10pt]
\\[-10pt]
\nonumber
&=&E\bigl[ E\bigl\{ E( Y \mid C_K=0,\bar{A}_{K-1}=
\bar{a}_{K-1},\bar{L}_{K-1},W ) \mid\\[-1pt]
&&\hspace*{62pt}{} C_K=0,
\bar{A}_{K-2}=\bar{a}_{K-2},\bar{L}_{K-2},W \bigr\}
\bigr]
\nonumber
\end{eqnarray}
by sequentially breaking up the expectations into nested conditional
expectations. This decomposition of the expectations is continued until
the outermost\vadjust{\goodbreak} expectation is only conditional on $W$.

In order to obtain an estimate of the parameter using this
decomposition, a~model must be fit for each level of conditioning,
beginning with the innermost expectation. To more easily refer to each
model fit, \citet{vdl:004} described the conditional models of the
counterfactuals iteratively. Let
\[
\bar{Q}_{K}=E( Y \mid C_K=0,\bar{A}_{K-1}=
\bar{a}_{K-1},\bar{L}_{K-1},W )
\]
be the outcome expectation conditional on the full history, for those
who followed the regime $\bar{a}$ and were fully observed. The fit $\bar
{Q}_{K}$ is obtained using a conditional modeling method. Then,
recursively define
\begin{eqnarray*}
\bar{Q}_{t}&=&E( \bar{Q}_{t+1} \mid C_{t}=0,
\bar{A}_{t-1}=\bar {a}_{t-1},\bar{L}_{t-1},W ), \qquad t=K-1,
\ldots,2,
\\
\bar{Q}_1&=&E( \bar{Q}_2 | W)
\end{eqnarray*}
for each successive nested expectation. The overbar in $\bar{Q}_t$
denotes a mean.

This alternative decomposition of the parameter can be used to compute
an estimate of the parameter of interest using the following algorithm.
It is done by producing model fits for each of the $\bar{Q}_t$'s,
obtaining predictions for each individual, and then taking a mean of
$\bar{Q}_1$ over all participants. Specifically, the estimation
algorithm proceeds as follows:
\begin{longlist}[1.]
\item[1.] First, model the outcome $Y$ given all of the covariate history,
for only those completely uncensored subjects with observed
breastfeeding regime $\bar{A}_{K-1}=\bar{a}_{K-1}$. This can be done
using logistic regression or any appropriate prediction method.
(Alternatively, a general conditional expectation conditional on $\bar
{A}_{K-1}$ can be fit using all uncensored subjects and then evaluated
at $\bar{a}_{K-1}$ in order to smooth over all observations.)
\item[2.] Then, using the model produced in (1), predict the conditional
outcome for all subjects (including those censored), resulting in the
fit $\bar{Q}_{K,n}$.

Then, iteratively for $t=K,\ldots,2$,
\item[3.]
Fit a model for $\bar{Q}_{t,n}$ from the previous step conditional on
covariates $\bar{L}_{t-1}$ using only subjects uncensored up until time
$t-1$ (i.e., subjects with $C_{t-1}=0$) with observed breastfeeding
status $\bar{A}_{t-2}=\bar{a}_{t-2}$. (Again, this model can be
alternatively fit using all uncensored subjects, conditioning on $\bar
{A}_{t-2}$, and then evaluating at $\bar{a}_{t-2}$.)
\item[4.] For all subjects, predict a new conditional outcome from this
last model, producing the fit $\bar{Q}_{t-1,n}$.
\end{longlist}
Repeat steps 3 and 4 for each time point (going backward in time) until
predictions $\bar{Q}_{1,n}$ are obtained for the outcome conditional on
only the baseline covariates, $W$. The parameter estimate is then
obtained by taking a mean of $\bar{Q}_{1,n}$ over all observations. As
in the previous G-computation method, variance estimates are computed
using bootstrap cluster resampling. Note that the above procedure does
not depend on the type or dimension of the variables $L_t$ and $W$, and
fits one model per time point (where there is an intervention or censoring).\vadjust{\goodbreak}

%

\subsection{Efficient estimation for longitudinal data}

Both G-computation algorithms described here require correct
specification of different decompositions of the underlying data
generating form. Alternatively, efficient semiparametric estimation
allows for root-$n$ consistent estimation with the added benefit of
double robustness [\citet{vdl:005,Tsiatis}]. Briefly, influence curves
are weighted score functions that contain all of the information about
the asymptotic variance of the related estimator. The \emph{efficient
influence curve} for a given parameter is the influence curve that
reaches the minimal variance bound. One possible way of obtaining
efficient semiparametric inference is to estimate the components of the
efficient influence curve and then use it as an estimating equation by
setting it equal to zero and solving for the target parameter.

Corresponding to the original G-computation factorization of the
likelihood, \citet{vdl:002} derived a representation of the efficient
influence curve for a longitudinal form with binary intermediate
variables. Similarly, \citet{Stitelman:001} modified the corresponding
theory for survival data. The alternative formulation for the efficient
influence curve was developed by \citet{Bang:001} and used by \citet
{vdl:004}, allowing for a general longitudinal form and much easier
estimation procedures for higher-dimensional or more complex
longitudinal data.

Let $\bar{g}_{t}, t=2,\ldots,K$, be the probability associated with
obtaining a given history of breastfeeding $\bar{a}$ up until time
$t-1$ and no censoring up until time-point $t$, conditional on the
observed history $\bar{L}_{t-1}$ and $W$. Specifically, let
%
\begin{eqnarray}\label{propensity}
&&\bar{g}_{t}(\bar{L}_{t-1},W)\nonumber\\
&&\qquad=\operatorname{Pr}(C_1=0 \mid W )
\nonumber
\\[-8pt]
\\[-8pt]
\nonumber
&&\qquad\quad{}\times\prod_{k=2}^{t} \bigl\{ \operatorname{Pr}(C_k
= 0 \mid\bar{A}_{k-1} = \bar{a}_{k-1}, C_{k-1} = 0,
\bar{L}_{k-1},W)\\
&&\hspace*{32pt}\qquad\quad{}\times\operatorname{Pr}(A_{k-1}=a_{k-1} \mid\bar{A}_{k-2}=
\bar{a}_{k-2}, C_{k-1}=0, \bar{L}_{k-1},W)
\bigr\}\nonumber
\end{eqnarray}
for $t=2,\ldots,K$, and where $A_0$ and $a_0$ are null sets. Further, let
$\bar{g}_{1}(W)=\operatorname{Pr}(C_1=0 \mid W )$ be the probability of being
uncensored at the first time point, conditional on baseline covariates,
$W$. These probabilities can be estimated using logistic regression,
for instance. As derived and explained for a general longitudinal
structure in~\citet{vdl:004}, the efficient influence curve $D(O)$ for
a fixed $\bar{a}$ can then be written recursively for the PROBIT data
as the sum of the components
%
\begin{eqnarray}\label{EIF}
D_{t}&=&\frac{I(\bar{A}_{t-1}=\bar{a}_{t-1},C_{t}=0)}{\bar{g}_{t}}(\bar {Q}_{t+1}-
\bar{Q}_{t}) \qquad\mbox{for } t=K,\ldots,2,\nonumber
\\
D_{1}&=&\frac{I(C_1=0)}{\bar{g}_{1}}(\bar{Q}_2-\bar{Q}_1)\quad
\mbox{and}
\\
D_0&=&(\bar{Q}_1-\psi_{\bar{a}}),
\nonumber
\end{eqnarray}
where $\bar{Q}_{K+1}=Y$ is defined for notational convenience (and the
dependencies of some components repressed). $I(\cdot)$ is an indicator function.

With each of the $\bar{g}_t$ and $\bar{Q}_t$ components estimated using
any given prediction method, the parameter $\psi_{\bar{a}}$ can be
estimated by setting the sum of the $K+1$ components equal to zero and
solving for $\hat{\psi}_{\bar{a}}$. In addition to being efficient,
such an estimator is double robust: it is consistent if either the
models for $\bar{Q}_t,t=1,\ldots,K$, or the models for $\bar
{g}_{t},t=1,\ldots,K$, contain the truth.

\subsection{TMLE using the alternative G-computation formulation}



The sequential G-computation method described in Section~\ref{altG} is
a substitution estimator because it is a function of a component of the
likelihood, specifically the nested conditional expectations, $\bar
{Q}_t$. The general TMLE procedure begins with some choice of
substitution estimator, but modifies this estimator by updating the
fits of the conditional expectations in order to produce a parameter
estimate that satisfies the equation of the efficient influence curve
set equal to zero. This parameter estimate is efficient and double
robust. The general TMLE procedure has been described previously, for
example, by \citet{vdl:001,Gruber:001,Rosenblum:001}.

Details regarding the construction of the sequential longitudinal
estimator are given by~\citet{vdl:004}. The first step in the TMLE
procedure is to fit the conditional densities $\{\bar{Q}_t,t=1,\ldots,K\}$
using a method of choice. For the update step, the logistic loss
function is chosen even for our case of an integer-valued outcome
(which is reduced to proportions by shifting and scaling the vector to
$[0,1]$) due to the boundedness properties of the inverse of its
canonical link function. The logistic loss becomes particularly
valuable when there is sparsity at certain levels of the covariates or
exposure [\citet{Gruber:001}].

The next step is to fluctuate each of the initial density estimates $\{
\bar{Q}_{t,n},t=K,\ldots,1\}$, starting at $t=K$, with respect to a new
parameter, $\varepsilon_t$. A subscript $n$ will be used to denote a
fitted value. The fluctuation function for each $\bar{Q}_{t}(\varepsilon
_t)$ can be described as
%
\begin{equation}
\operatorname{logit}\bar{Q}^1_{t}(\varepsilon_t)=
\operatorname{logit}\bar {Q}_{t}+\varepsilon_t G_t,\qquad
t=1,\ldots,K, \label{flux2} 
\end{equation}
for some expression $G_t$.
Again letting $\bar{Q}_{K+1}=Y$, the estimate for $\varepsilon_t$ is
found by minimizing the empirical mean of the logistic loss function
%
\begin{equation}
\label{loss2}\mathcal{L}\bigl\{\bar{Q}^{1}_{t}(
\varepsilon_t)\bigr\}=-\bigl[\bar {Q}_{t+1}\log\bigl\{
\bar{Q}^{1}_{t}(\varepsilon_t)\bigr\}+(1-
\bar{Q}_{t+1})\log\bigl\{ 1-\bar{Q}^{1}_{t}(
\varepsilon_t)\bigr\}\bigr],
\end{equation}
which is equivalent to solving the empirical mean score (or derivative
of the loss function) at zero. This requires that the function $G_t$ be
defined and estimated.

According to the general TMLE procedure, the above fluctuation
function in equation~(\ref{flux2}) is required to satisfy two
conditions: (1) the fluctuation function must reduce to the original
density when $\varepsilon_t=0$, and (2) the derivative with respect to
$\varepsilon_t$ of the loss function at $\varepsilon_t=0$ must linearly span
the efficient influence curve. The first condition is clearly satisfied
when $\varepsilon_t=0$. Taking the derivative of the loss function in
equation~(\ref{loss2}) with respect to $\varepsilon_t$ gives
\[
\frac{d \mathcal{L}(\bar{Q}^{1}_{t,n}(\varepsilon_t))}{d\varepsilon_t}
\bigg\vert_{\varepsilon_t=0}=G_t\times(
\bar{Q}_{t+1}-\bar{Q}_{t}),\qquad t=1,\ldots,K.
\]
Therefore, the score spans the efficient influence curve when $G_t$ is
defined as
\[
G_t(C_t,\bar{A}_{t-1},\bar{L}_{t-1},W)=
\frac{I(C_t=0,\bar{A}_{t-1}=\bar
{a}_{t-1})}{\bar{g}_{t}}.
\]
The covariate $G_t$ is often described as ``clever'' because it allows
the score to span the efficient influence curve.

The update step is carried out by minimizing the empirical mean of the
loss function, $\mathcal{L}\{\bar{Q}^{1}_{t,n}(\varepsilon_t)\}$, with
respect to $\varepsilon_t$. This is equivalent to running the logistic
regression in equation~(\ref{flux2}): no intercept, with offset
logit$(\bar{Q}_{t,n})$ and unique covariate $G_t(C_t,\bar{A}_{t-1},\bar
{L}_{t-1},W)$. Let $\hat\varepsilon_t$ be the estimate of the coefficient
for $G_t$, which is the maximum likelihood estimate (or, equivalently,
the minimum loss-based estimate) for $\varepsilon_t$.

Once all of the densities have been updated to give $\{\bar
{Q}^{1}_{t,n},t=K,\ldots,1\}$, the parameter $\psi_{\bar{a}}$ is estimated
as the mean of $\bar{Q}^{1}_{1,n}$ over all subjects, that is, $\hat
{\psi}_{\bar{a}}=\frac{1}{n}\sum_i \bar{Q}^{1}_{1,n}(W=w_i)$ (where
$w_i$ is the observed baseline vector for subject $i$).

This TMLE is double robust: it is consistent if either the models for
$\bar{Q}_t,t=1,\ldots,K$, or the models for $\bar{g}_{t},t=1,\ldots,K$,
contain the truth. In addition, because of the usage of the logistic
loss function and the corresponding fluctuation function in
equation~(\ref{flux2}), the parameter estimates are bounded, regardless
of the size of the weights, $\bar{g}_{t}^{-1}$. This makes TMLE robust
to certain kinds of data sparsity that cause large weights.
A comparison of the fundamental qualities of the G-computation
estimators, TMLE and IPTW, can be found in Table~\ref{compare}.
%
\begin{table}
\caption{Comparison of methods}\label{compare}
\begin{tabular*}{\textwidth}{@{\extracolsep{\fill}}lcccc@{}}
\hline
 & \textbf{Required for} & \textbf{Robust to} & \textbf{Variance} & \textbf{Respects parameter}\\
\textbf{Method}& \textbf{consistency} & \textbf{data sparsity} & \textbf{estimate}& \textbf{boundaries}\\
\hline
G-comp.& CE & $\checkmark$ & BS & $\checkmark$ \\
G-comp. seq.& NE & $\checkmark$ & BS & $\checkmark$ \\
IPTW & propensity & $\times^*$ & EIC/BS & $\times$\\
TMLE & propensity or NE & $\checkmark$ & EIC & $\checkmark$ \\
\hline
\end{tabular*}
\tabnotetext[]{}{CE: conditional expectations; NE: nested expectations;
BS: bootstrap; EIF: efficient influence curve; propensity: the
conditional probabilities of intervention (e.g., breastfeeding) and
censoring. *Improvement under weight stabilization.}
\end{table}

\subsubsection{TMLE procedure for the PROBIT data}\label{PROBITproc}

We observed the following procedure in our estimation of the parameter
$\psi_{\bar{a}}$, for a given breastfeeding regimen $\bar{a}$. As
described above, our interpretation of the structure of the PROBIT data
set is $O=(W,C_1,L_1,A_1,C_2,L_2,\ldots,A_{5},C_{6},Y)$. There are six
intervention nodes: censoring can occur at any of them and
breastfeeding status is assessed at $t=1,\ldots,5$. All subjects are
initially breastfeeding, so breastfeeding regimen is equivalent to the
total duration of breastfeeding. If a subject has been censored, impute
their missing $L_t$ and $Y$ variables with zero values:

\begin{enumerate}[1.]
\item[1.] Fit models predicting breastfeeding and censoring (resp.)
at each time point, conditional on all previous history. For each
model, compute a predicted probability for each subject conditional on
$\bar{A}_t=\bar{a}_t$ and $C_t=0$.
\begin{itemize}
\item Given the monotone nature of breastfeeding, if $\bar
{a}=(1,0,0,0,0)$, for instance, the predicted probability of \emph{not}
breastfeeding at time 3 will be one for all participants, since it is
conditional on stopping before time 2.
\end{itemize}

\item[2.] Using the predictions from step 1, calculate the propensity score
$\bar{g}_{t,n}$ from equation~(\ref{propensity}) for each subject.

\item[3.] Set $\bar{Q}_{7,n}=Y$, where $Y$ is rescaled to $[0,1]$. Then, for
$t=6,\ldots,1$,
\begin{itemize}
\item For the subset of subjects with $\bar{A}_{t-1}=\bar{a}_{t-1}$ and
$C_t=0$, fit a model for $E(\bar{Q}_{t+1,n} \mid\bar{L}_{t-1})$. Using
this model, predict the conditional outcome for all subjects and let
this vector be denoted $\bar{Q}_{t,n}$.
\item Construct the ``clever covariate'' $G_t(C_t,\bar{A}_{t-1},\bar
{L}_{t-1},W)=I( C_t=0,\break \bar{A}_{t-1}=\bar{a}_{t-1})/\bar{g}_{t,n}$.
\item Update the expectation by running a no-intercept logistic
regression with outcome $\bar{Q}_{t+1,n}$, the fit $\operatorname{logit}(\bar
{Q}_{t,n})$ as an offset and clever covariate $G_t$ as the unique
covariate. Let $\hat\varepsilon_t$ be the estimated coefficient of $G_t$.
\item Update the fit of $\bar{Q}_t$ by setting
\[
\bar{Q}^{1}_{t,n}=\operatorname{expit}\bigl\{\operatorname{logit}(
\bar{Q}_{t,n})+\hat\varepsilon _tG_t(
\bar{A}_{t-1}=\bar{a}_{t-1},C_t=0,
\bar{L}_{t-1})\bigr\}
\]
and then obtain a predicted value of $\bar{Q}^{1}_{t,n}$ for all
subjects.\\
Note that the model for $\bar{Q}_1$ is modeled using only subjects
with $C_1=0$. The resulting fit $\bar{Q}_{1,n}$ is only conditional on
$W$ and is estimated for all subjects.
\end{itemize}
\item[4.] Having fit $\bar{Q}^{1}_{1,n}$ for each subject, take the mean.
Rescale the mean (do the inverse of the original scaling of $Y$). This
is the TMLE for $\psi_{\bar{a}}$.
\end{enumerate}

The standard errors can be calculated using a sandwich estimator,
which uses the influence curve to approximate the asymptotic variance.
First, the value of the influence curve $D(O)$ is estimated for each
subject. The clusters $Z_m$ are indexed by $m=1,\ldots,M$. Let $\rho
_m=E(D_iD_j)$ for two elements in the cluster $Z_m$ and let $\sigma
^2_m=\operatorname{Var}(D_i)=E(D_i^2)$ be the common variance for subjects in cluster
$Z_m$. Assuming independence between the clusters and common variance
for elements in a cluster, the large sample variance of the estimator
is approximated using
\begin{eqnarray*}
\sigma^2&=&1/n^2E \Biggl(\sum_{i=1}^n
D_i \Biggr)^2 = \frac
{1}{n^2} \sum
_{m=1}^M\sum_{i,j\in Z_m}
E(D_iD_j)I(i\neq j) + E\bigl(D_i^2
\bigr)I(i=j)
\\
&=&\frac{1}{n^2} \sum_{m=1}^M
n_m(n_m-1) \rho_m + n_m
\sigma_m^2,
\end{eqnarray*}
where $n_m$ is the size of cluster $Z_m$. The supplemental article~\citet
{Sup:SeqTMLE} contains details about the form of the influence curve
under clustering.
The expectations can be estimated by taking the empirical covariance
and variance within each of the clusters. Confidence intervals are
calculated assuming Normality of the estimator, using the estimate plus
and minus 1.96 times the estimated standard error.

\section{Analysis of the PROBIT}\label{sec4}

The PROBIT data were analyzed by both G-computation methods; TMLE with
parametric modeling of the sequential conditional means and conditional
probabilities of breastfeeding and censoring (logistic main terms
regression for binary breastfeeding status and censoring, and for the
outcome shifted and scaled to $[0,1]$); TMLE with Super Learner to model
the conditional expectations and probabilities; and a stabilized IPTW
estimator. All models were implemented directly in R Statistical
Software [\citet{R}] with the exception of Super Learner which we fit
using the R library \texttt{SuperLearner} [\citet{SL}]. Super Learner
calculates predictions using each method in a library, and then
estimates the ideal combination of these results based on the $k$-fold
cross-validated error. The library we chose included main terms
logistic regression, generalized additive modeling [\citet{gam}], the
mean estimate, a nearest neighbor algorithm [\citet{Ipred}],
multivariate adaptive regression spline models [\citet{Earth}] and a
stepwise AIC procedure [\texttt{stepAIC} from \citet{MASS}].

A stabilized IPTW estimator was computed by obtaining the solution of
the empirical mean of
\[
\bigl(Y-\hat{\psi}_{\bar{a}}^{\mathrm{IPTW}}\bigr)\frac{I(\bar{A}_{5}=\bar{a},C_6=0)({1}/{n})\sum\bar{g}_{6,n} }{\bar{g}_{6,n}}
\]
set equal to zero. To be consistent, IPTW relies on correct modeling of
the breastfeeding and censoring probabilities in $\bar{g}_6$. IPTW was
implemented using logistic regressions to fit each of these conditional
probabilities.

\begin{table}
\caption{Differences in marginal expected number of infections under
different breastfeeding durations}\label{results2}
\begin{tabular*}{\textwidth}{@{\extracolsep{\fill}}ld{2.3}d{1.3}c@{}}
\hline
\textbf{Method} & \multicolumn{1}{c}{\textbf{Estimate}} & \multicolumn{1}{c}{\textbf{S.E.}}& \textbf{95$\%$ C.I.}\\
\hline
\multicolumn{4}{c}{\emph{3--6 months vs 1--2 months}}\\
G-comp. (likelihood)& -0.032& 0.008& $(-0.046,-0.019)$\\
G-comp. (sequential)& -0.039& 0.013& $(-0.062,-0.016)$\\
IPTW& -0.021 & 0.011 & $(-0.042,0.000)$\\
Parametric TMLE& -0.027 & 0.010 & $(-0.045,-0.008)$\\
TMLE with SL& -0.039 & 0.010 & $(-0.058,-0.020)$\\[3pt]
\multicolumn{4}{c}{\emph{9$+$ months vs 3--6 months}}\\
G-comp. (likelihood)& -0.013& 0.004& $(-0.020,-0.005)$\\
G-comp. (sequential)& -0.014& 0.013& $(-0.027,0.004)$\\
IPTW& -0.013 & 0.010 & $(-0.032,0.007)$\\
Parametric TMLE& -0.021 & 0.013 & $(-0.047,0.004)$\\
TMLE with SL& -0.024 & 0.007 & $(-0.038,-0.010)$\\[3pt]
\multicolumn{4}{c}{\emph{9$+$ months vs 1--2 months}}\\
G-comp. (likelihood)& -0.045& 0.010& $(-0.065,-0.027)$\\
G-comp. (sequential)& -0.053& 0.018& $(-0.084,-0.020)$\\
IPTW& -0.034 & 0.014 & $(-0.061,-0.007)$\\
Parametric TMLE& -0.048 & 0.018 & $(-0.084,-0.012)$\\
TMLE with SL& -0.063 & 0.013 & $(-0.088,-0.038)$\\
\hline
\end{tabular*}
\tabnotetext[]{}{G-comp.: G-computation, using both methods described in
the text, likelihood in Section~\ref{Gcomp2} and sequential in
Section~\ref{altG}; TMLE: targeted maximum likelihood estimation; SL:
Super Learner; IPTW: inverse probability of treatment weighting (stabilized).}
\end{table}

The standard errors for all methods except the G-computations were
calculated using the sandwich estimator, adjusting for clustering as
described in Section~\ref{PROBITproc}. The standard errors for the
G-computation methods were estimated using pairs cluster
bootstrap~[\citet{Cameron:001}] by resampling the 31 clusters with
replacement, repeating 200 times, recalculating the estimates, and
taking the standard error of the estimates. Confidence intervals were
calculated by taking the 2.5th and 97.5th quantiles of the resampled estimates.

Both G-computations were found to be sensitive to modeling choices when
fitting the conditional expectations. In particular, we implemented
both G-computations with Poisson regressions and with logistic
regressions using a rescaled outcome. For the standard G-computation,
both parametric specifications produced very similar point estimates,
but the Poisson model was found to be highly unstable through the
cluster bootstrapping while the logistic model was more stable. For the
sequential G-computation, the Poisson model produced uninterpretable
point estimates that deviated substantially from the other models,
while the point estimates of the logistic model conformed more or less
to the other results. Only the logistic results are therefore presented
in the table.

The estimates of three comparisons of interest are presented in
Table~\ref{results2}. The first parameter of interest is the difference
between the marginal expected number of infections (in the first year
or life) for infants who were breastfed for between 3 and 6 months
compared to infants who were breastfed for between 1 and 2 months. The
second parameter compares infants who were breastfed for greater than 9
months to those breastfed for 3 to 6 months. The third parameter
compares greater than 9 months to between 1 and 2. The table presents
the estimates, standard errors and 95\% confidence intervals for each
parameter of interest as calculated by each method.\looseness=1

All of the methods estimated a negative parameter value for the
difference, corresponding with the interpretation that longer durations
of breastfeeding reduce the expected number of gastrointestinal
infections. TMLE with Super Learner and likelihood G-computation found
a statistically significant difference for each comparison. Only IPTW
found an insignificant estimate for the first comparison. Sequential
G-computation, IPTW and parametric TMLE found an insignificant estimate
for the second comparison. All methods determined that there is a true
difference between the marginal mean infection counts for breastfeeding
for over nine months versus between one and two months.

The estimates of the difference parameters varies substantially between
methods. In two of the comparisons, TMLE with Super Learner produced
higher estimates than all of the other methods (almost twice the size
of the smallest estimates). IPTW gave the smallest estimates of the
differences. Likelihood G-computation consistently produced the
smallest standard errors and TMLE with Super Learner produced the
second smallest.

\subsection{The validity of a causal interpretation}\label{causalassumptions}

A causal interpretation of the analysis of the PROBIT data requires
several important but untestable assumptions, including the sequential
randomization assumption. In other words, all confounders are assumed
to have been measured and included in $W$, including all prognostic
factors of infection that also predict censoring. The complexities of
the substantive matter make it challenging to believe that we
identified all the common causes of breastfeeding cessation and
infections [\citet{Kramer:003}]. However, we argue that by controlling
for cluster as a baseline variable, much of this confounding effect may
have been alleviated (this is investigated in Section~\ref{simstudy}).

In addition, we must assume no interference between study units
(mother/\break infant pairs) and that only one version of the treatment (i.e.,
breastfeeding) is applied to all units [together referred to as the
stable unit treatment variable assumption, or SUTVA; \citet{Rubin:002}].
The assumption of no interference requires that the breastfeeding
status of one mother does not influence the outcome of another's child.
We believe this to be very plausible because mothers spent short
periods of time in the hospital which limited their interaction. For
the second assumption, due to the discretization of the study design,
different durations of breastfeeding are grouped together. We must
assume that it does not matter when a mother ceases to breastfeed
within an interval.

\section{Simulation study}\label{simstudy}

A simulation study was performed where data were generated as a
simplified version of the PROBIT data set. Five hundred subjects were
generated in each of 31 clusters. The baseline covariates $W$ and $U$
were generated as Gaussian variables with cluster-specific means drawn
from separate Gaussian distributions. The time-dependent variables
$(C_1,L_1,A_1,C_2,L_2,A_2,C_3, L_3)$ were generated independently for
each subject conditional on the subject's history, including baseline
variables $W$ and $U$ (and not otherwise clustered). Binary variables
$A_t, t=1,2$, indicate continued breastfeeding, $C_t, t=1,2,3$, are
censoring indicators, and $L_t, t=1,2,3$, indicate the presence of
infections. The outcome $Y=\sum_{t=1}^3 L_t$ is a count variable.
Breastfeeding status was generated as conditional on the baseline
variables and immediate preceding covariates at every time point. In
particular, breastfeeding was specifically made to be less likely to
continue when infection was indicated at the current time point.
Breastfeeding (like censoring) is a monotone process, and so $A_2=1$ is
only possible if $A_1=1$. The probability of censoring was conditional
on baseline covariates and most recent infection status; censoring was
less likely if breastfeeding continued at the previous time point and
more likely if an infection occurred at the previous time point.
Infections were generated conditional on baseline variables and
breastfeeding for the past two visits, so that longer duration of
breastfeeding decreased the probability of infection. The strengths of
the associations between exposure/censoring and intermediate infections
were designed to reflect the true PROBIT results. Details of the data
generation can be found in the supplemental article~\citet{Sup:SeqTMLE}.

The parameter $\psi_{\bar{a}}=E(Y_{\bar{a}})$ was estimated for $\bar
{a}=(0,0)$ and $\bar{a}=(1,1)$. The parameter of interest, reflecting
the first parameter of interest in the PROBIT study, was $\delta=\psi
_{(1,1)} - \psi_{(0,0)}$.

A concern we had during the planning of the PROBIT study was that we
may be missing some important confounders of the effect of
breastfeeding on infection. Therefore, we attempt to explore this issue
in the simulation study by omitting the variable $U$ from the modeling.
In a second modeling scenario, we illustrate how adjusting for the
cluster like a baseline confounder can successfully adjust for
unmeasured confounding that is characterized by the cluster itself. In
addition, we test the scenario where $U$ is included in the modeling so
that the results could be compared. Finally, we test a scenario where
we suppose that the analyst is given transformed versions of $W$ and~$U$ [using two of the transformations in~\citet{Kang:001}] and the
models are run using these transformed variables.

\begin{table}
\caption{Difference between marginal expected outcomes, by scenario.
True value${} = -0.030$}\label{simresults2}
\begin{tabular*}{\textwidth}{@{\extracolsep{\fill}}ld{2.3}d{4.0}ccd{3.0}@{}}
\hline
\textbf{Method} & \multicolumn{1}{c}{$\bolds{\hat\delta}$} &
\multicolumn{1}{c}{\textbf{\% bias}} &
\multicolumn{1}{c}{$\bolds{\operatorname{SE}(\hat\delta)}$} &
\multicolumn{1}{c}{$\bolds{\operatorname{rMSE}(\hat\delta)}$} &
\multicolumn{1}{c@{}}{\textbf{Coverage}$^\mathbf{a}$}\\
\hline
&\multicolumn{5}{c}{\emph{Unmeasured confounder}}\\
G-comp. (likelihood)& -0.060&-99&0.017&0.035&49\\
G-comp. (sequential)& -0.062&-105&0.018&0.037&44\\
IPTW& -0.054&-77&0.021&0.023&100\\
Parametric TMLE& -0.058&-90&0.017&0.027&63\\
SL TMLE&-0.054&-79&0.019&0.024&78\\[3pt]
&\multicolumn{5}{c}{\emph{Unmeasured confounder, adjusting for
cluster}}\\
G-comp. (likelihood)&-0.033 &-11&0.008&0.009&92\\
G-comp. (sequential)& -0.035&-16&0.009&0.011&94\\
IPTW& -0.032&-6&0.010&0.009&94\\
Parametric TMLE& -0.032&-7&0.009&0.009&94\\
SL TMLE& -0.030&1&0.008&0.009&90\\[3pt]
&\multicolumn{5}{c}{\emph{Adjusting for all confounders}}\\
G-comp. (likelihood)& -0.032&-4&0.008&0.009&91\\
G-comp. (sequential)& -0.034&-12&0.018&0.010&43\\
IPTW& -0.031&-1&0.010&0.009&93\\
Parametric TMLE& -0.031&-1&0.009&0.009&92\\
SL TMLE& -0.029&5&0.009&0.010&88\\[3pt]
&\multicolumn{5}{c}{\emph{Transformed confounders}}\\
G-comp. (likelihood)& -0.068& -125&0.017 &0.042 &29\\
G-comp. (sequential)& -0.075& -147& 0.023&0.050 &20\\
IPTW& -0.062& -106&0.109 &0.125 &55\\
Parametric TMLE&-0.067 &-121 &0.041 &0.045 &36\\
SL TMLE&-0.033&-9&0.032&0.013&95\\
\hline
\end{tabular*}
\tabnotetext[]{}{SE($\delta$): the average standard error is the
square-root of the mean of the variances, with each variance calculated
using the influence curve for TMLE and IPTW and the nonparametric
boostrap$^\mathrm{b}$ for G-comp. (likelihood) and G-comp. (sequential); rMSE:
root mean squared error calculated over the simulated data sets;
Coverage: mean coverage; TMLE: targeted maximum likelihood estimator;
G-comp.: G-computation; IPTW: (stabilized) inverse probability of
treatment weighting.
$^\mathrm{a}$The estimated coverage is the $\%$ of data sets where the
true value falls between (i) the estimate plus and minus 1.96 times the
standard error of the estimate for TMLE and IPTW or (ii) the 2.5th and
97.5th bootstrap percentiles for the G-computation methods; $^\mathrm{b}$The
bootstrap standard error was computed using 200 resamples from the data
set of size $n=15\mbox{,}500$.}
\end{table}

One thousand data sets of $500\times31=15\mbox{,}500$ observations were
generated. Under each of the four modeling scenarios (unmeasured $U$,
adjusting for cluster, adjusting for $U$ and transformed confounders),
the performance of the TMLE was compared to G-computation, the
sequential formulation of the G-computation formula and a stabilized
IPTW estimator. TMLE was implemented in two ways: with main terms
logistic regressions to estimate all probabilities and with Super
Leaner, using only main terms logistic regression and a nearest
neighbors algorithm in its library (a small subset of the library used
in the PROBIT analysis). Standard errors were computed using influence
curve inference where available and nonparametric bootstrap resampling
otherwise (details in the footnote of Table~\ref{simresults2}). Due to
the way the data were generated, the sequential G-computation was
always incorrectly specified (in the model form), as were the outcome
models for the TMLE.

As a small departure from the real data, the simulated data allowed
only one infection at each time interval (as opposed to more than one
event). The G-computation used the information that the outcome was a
sum of the first two binary infection variables and the additional
binary variable, $L_3$, measured at time $t=3$. Thus, $Y=\sum_{t=1}^2
L_t + L_3$, so that the G-computation simplified to the empirical mean of
\begin{eqnarray*}
&& \sum_{l_1=\{0,1\}}\cdots\sum_{l_K=\{0,1\}}
\Biggl[ \Biggl\{\sum_{t=1}^2
L_t + E(L_3\mid C_3=0,\bar{A}_2=
\bar{a}_2,\bar{L}_2=\bar{l}_2,W) \Biggr\}\\
&&\hspace*{60pt}\qquad{}\times \bigl\{p(L_2=l_2\mid C_2=0,
\bar{A}_{1}=\bar{a}_{1}, \bar {L}_{1}=
\bar{l}_{1},W) \bigr\}\\
&&\hspace*{190pt}{}\times p(L_1=l_1|C_1=0,W)
\Biggr].
\end{eqnarray*}

Note that using the information regarding the number of infections at
each time interval for the PROBIT data analysis would have required
fitting multinomial models in the likelihood G-computation. With so few
subjects having more than one infection at any given time, we did not
feel that substantial information could be added by increasing the
complexity of the model for the applied example using a similar approach.

\subsection{Simulation results}

The results of each of the models under each modeling scenario are
displayed in Table~\ref{simresults2}. With an unmeasured confounder
related to cluster, both G-computation models performed the most poorly
in terms of bias, root mean-squared error (rMSE) and coverage. TMLE
produced an improvement in these measures, and adding Super Learner
improved all measures of performance except for the standard error.
IPTW had the lowest bias, but higher standard errors, resulting in
overcoverage. When cluster was used as a surrogate for the unmeasured
confounder, all of the methods produced results with much lower bias
and standard errors. When all confounders were measured and adjusted
for, G-computation, parametric TMLE and IPTW all had a reduction in
bias compared to the previous scenario and performed ideally, despite
parametric TMLE being model-misspecified in the outcome models. TMLE
with Super Learner produced slight undercoverage. The sequential
G-computation was model-misspecified and produced high bias and
standard error, leading to poor coverage (since it is not double
robust). When the confounders were transformed, all of the parametric
models were incorrectly specified, leading to high bias and low
coverage. Among the parametric models, IPTW had the lowest bias and the
highest root mean-squared error. TMLE with Super Learner (using only
one data adaptive algorithm in its library) was essentially unbiased
with ideal coverage.

\section{Discussion}\label{sec6}

In this article we applied five different causal methods to the PROBIT
data to obtain estimates of the differences in the marginal expected
number of infection counts for different breastfeeding durations. All
methods agreed that extending the duration of breastfeeding
significantly lowers the expected number of gastrointestinal
infections. TMLE with Super Learner produced much larger effect
estimates, for example, its estimate was almost double the IPTW
estimate for the comparison between 1--2 and 9$+$ months of breastfeeding.
This represents a clinically important difference in the estimated
effect. Super Learner also reduced the higher standard error of the
TMLE procedure to a level comparable to that of the G-computation
(which is an efficient parametric estimator).


Using the mean estimate from TMLE with Super Learner, altering the
breastfeeding durations of 16 mothers from between one and two months
to over nine months will avoid one infant infection (i.e., the Number
Needed to Treat or NNT) on average in this population. This can roughly
be compared with the intention-to-treat result in the original PROBIT
study [\citet{Kramer:001}], where they obtained a NNT of 24 for the
presence of \emph{any} gastrointestinal infection over the first year
when contrasting subjects who did and did not receive the breastfeeding
intervention. We have therefore shown that breastfeeding itself might
have a larger impact on childhood infections than suggested by the
original PROBIT analysis.


In the simulation study we generated baseline confounders from a
distribution with a cluster-specific mean. The simulation results
demonstrated that bias (and inflated standard error) incurred by
cluster-specific unmeasured confounders can be adjusted for using the
cluster indicators themselves as baseline covariates. We also showed
that under the plausible scenario of being given transformed versions
of the confounders, only TMLE with Super Learner was able to unbiasedly
estimate the parameter of interest.

TMLE is a double-robust method, as it only requires correct
specification of the conditional probabilities of the intervention
(here, breastfeeding and censoring) or of the nested conditional
expectations of the outcome (the $\bar{Q}_t$'s) to be consistent.
Contrastingly, IPTW relies on correct specification of the
probabilities of the intervention, and the G-computations rely on
correct specification of the outcome models. When the probabilities of
intervention are modeled in the same way for IPTW and TMLE, in absence
of data sparsity, and when the outcome models are incorrectly
specified, these two methods are expected to perform similarly (as seen
in the simulation study and possibly in the PROBIT results). In many
other contexts, advantages of longitudinal TMLE over IPTW and
G-computation have been established through simulation study in \citet
{vdl:004}, \citet{Petersen:001}, \citet{Stitelman:001}, and \citet
{Schnitzer:001}.

It is important to note that for longitudinal data with time-dependent
confounding, there may not exist a data generating distribution that
corresponds to the way the outcome is modeled in the TMLE (i.e., in the
sequential G-computation). Therefore, we recommend that data adaptive
methods like Super Learner always be used with TMLE in the longitudinal
setting. Because TMLE with Super Learner is arguably the most reliable
estimator (assessed through theory and simulation studies), we have
reason to believe that the magnitude of the effect of breastfeeding is
actually larger than suggested by the methods that use parametric
modeling and larger than the effect reported in the original PROBIT analysis.

\section*{Acknowledgments} The authors acknowledge the usage of
Consortium Laval, Universit\'e du Qu\'ebec, McGill and Eastern Qu\'ebec
computing resources. In addition, the authors would like to thank
Michael Kramer for unrestricted access to the PROBIT data set. The
PROBIT was supported by grants from the Thrasher Research Fund, the
National Health Research and Development Program (Health Canada),
UNICEF, the European Regional Office of WHO and the CIHR.


\begin{supplement}[id=suppA]
\stitle{The efficient influence curve for clustered data and data
generation for the simulation study}
\slink[doi]{10.1214/14-AOAS727SUPP} 
\sdatatype{.pdf}
\sfilename{aoas727\_supp.pdf}
\sdescription{Derivation
of the efficient influence curve used in the TMLE analysis. Full
description (with R code) of the data generation used in the simulation study.}
\end{supplement}

%

\printaddresses

\end{document}